# Commuting network model: going to the bulk

F. Gargiulo[1], M. Lenormand[2], S. Huet[2], O. Baqueiro Espinosa[3]
[1]INED, 133 boulevard Davout, 75020, Paris, France
[2]LISC, Cemagref, BP 50085, 63172 Aubière, France
[3]IAMO, Theodor-Lieser-Str.2, D-06120 Halle (Saale), Germany

## Abstract

The influence of commuting in socio-economic dynamics increases constantly. Analysing and modelling the networks formed by commuters to help decision-making regarding the land-use has become crucial. This paper presents a simple spatial interaction simulated model with only one parameter. The proposed algorithm considers each individual who wants to commute, starting from their living place to all their workplaces. It decides where the location of the workplace following the classical rule inspired from the gravity law consisting in a compromise between the job offers and the distance to the jobs. The further away the job offer is, the more important it must be in order to be considered. Inversely, only the quantity of offers is important for the decision when these offers are close. The paper also presents a comparative analysis of the structure of the commuting networks of the four European regions to which we apply our model. The model is calibrated and validated on these regions. Results from the analysis shows that the model is very efficient in reproducing most of the statistical properties of the network given by the data sources.

## 1. Introduction

For two decades, not only the number of commuters, but also the average distance travelled by workers has been increasing in most European countries. This makes commuting a fundamental phenomenon to understand socio-economic macrostructures. A precise description of the commuting patterns has a central role in many applied questions: from the studies on traffic and the planning of infrastructures (Ortuzar and Willusem 2001) to the diffusion of epidemics (Balcan et al. 2009) or large demographic simulations (Huet and Deffuant 2011). The literature on this argument is abundant, both from the point of view of the analysis of the structures and from the point of view of the models.

Many recent papers have adopted an approach based on network theory. An interesting and complete analysis of the commuting structures from this point of view has been introduced for example in (De Montis et al. 2007; De Montis et al. 2010). In this framework, most importantly concerning the modelling issues, the question about the commuting networks is set in a larger conceptual category: spatially constrained network structures. This kind of analysis concerns not only commuting, but all the situations where the geography has a significant role: from the reconstruction of migrant patterns (Lemercier and Rosental 2008) to the analysis of the internet at autonomous system (AS) level (Pastor-Satorras and Vespignani 2004), to the airline network structure (Barrat et al. 2004). A particularly important study in this context is, for example, (Barrat et al. 2005) where the concept of "preferential attachment" (Barabasi and Albert 1999) is adapted in order to take into account not only the strength of a node given by its current in-degree, but also the spatial constraint included in the journey-to-work network.

A more traditional approach to the commuting structures is based on the so-called gravity law models (Haynes and Fotheringham 1988). The term gravity law is a metaphor from classical physics. We can imagine that, as it happens in gravitation, the interaction between two municipalities depends proportionally on a parameter, for example the size of the municipality (equivalent to mass in the gravitational law), and in inverse proportion with some power law of the distance[1]. We notice that the concept of "distance" can be connected to something that is not precisely a real geographical/spatial category: it can be a travelling time, a topological distance on a network, but also a "social" distance. Imagine for example the cases of border cities where a different language is spoken on the other side. There are many versions of this law studied to better fit the data. Sometimes the inverse proportion with the distance ($d_{ij}$) is replaced by an exponential decay. In general, we can speak of a deterrence function $h(d_{ij})$. The literature generally agrees that an exponential specification appears better for modelling commuting. However, in some applications, a power law decay often seems to be a better fit (De Montis et al. 2007; De Montis et al. 2010; Reggiani and Vinciguerra 2007). Some studies propose a combined form of the two (Ortuzar and Willusem 2001) or a different form (de Vries et al. 2009) in order to better fit the empirical data.

---
[1] Let's notice that the power is 2 for gravitation.

We can consider different proportionality parameters $M_i$, $N_j$ (respectively for the origin and destination municipality): the size of the municipalities, the number of active people, the total number of out-commuters ($R_i$) and of in-commuters ($Q_i$). To each flow of commuters ($T_{ij}$) the probability $p_{ij}$, extracted from a gravity law, is associated:

$$p_{ij} = \frac{f(M_i)g(N_j)h(d_{ij})}{\sum_{i,j} f(M_i)g(N_j)h(d_{ij})}$$

Using these probabilities, it is possible to determine the traffic between each couple of municipalities with different methods (e.g. IPF, multinomial models, etc.). We notice that the functions $f(M_i), g(N_i)$ and $h(d_{ij})$ can assume any possible shape. The main point is that in general, whatever the indicator for $M_i$ and $N_i$, and whatever the kind of distance, this type of procedure requires at least three different parameters to calibrate independently (one for each function).

The most common form of a spatial interaction model inspired from the gravity law to model the commuting network is the so called "doubly-constrained" model (Wilson 1998) (Choukroun 1975). It predicts the number of journeys-to-work between any pair of origin-destination zones:

$$T_{ij} = A_i B_j R_i Q_j h(d_{ij})$$

where:

$$A_i = \frac{1}{\sum_j B_j Q_j f(\beta, c_{ij})} \text{ and } B_j = \frac{1}{\sum_i A_i R_i f(\beta, c_{ij})}$$

The balancing factors $A_i$ and $B_j$ ensure that the $T_{ij}$ table is consistent with the exogenous rows and column totals: $\sum_j T_{ij} = R_i$ and $\sum_i T_{ij} = Q_j$. These balancing factor, plus a distance parameter $\beta$, implicit in the function $h(d_{ij})$ have to be calibrated.

From the point of view of modelling, this procedure is an optimization method that associates a destination municipality to each commuter, maximising the agreement with the aggregate system setup. Any particular microstate will be associated with a macrostate, which is simply the number of trips from an origin to a destination. A macrostate is feasible if it reproduces known properties referred to as system states (for example the total number of travellers). Estimating the solution of the model consists in finding the macrostates, maximizing a chosen distance function of the considered macrostate to the observed data, among the feasible macrostates (Bernstein 2003).

A lot of improvements have been proposed based on this doubly-constrained model. In (Fotheringham 1981), a competing destination model is introduced in order to improve the spatial structure of the generated network. (Fik and Mulligan 1990) extend this competing model to measure the accessibility of a destination relatively to destinations of the same hierarchical order in the system of central places (founded on the Central Place Theory). They also incorporate a measure that relates to the number of intervening opportunities from the living place $i$ to the attractive force $j$. These intervening opportunities are the potential destinations within a distance smaller than $d_{ij}$. To go beyond the gravity law models' weaknesses, some authors developed an approach founded on the network paradigm (Thorsen et al. 1999) (Gitlesen et al. 2010). This kind of procedure has the disadvantage of increasing the number of parameters, which is what we want to avoid.

Indeed, our research takes place in the framework of the European project PRIMA[2]. The microsimulation model developed within the PRIMA project simulates the dynamics of the population living in the European rural municipalities. Therefore one of our main focuses is the commuting structures in the rural areas of our regional case study regions. These structures have to be analysed and reproduced in the microsimulation model which aims to help the decision regarding land-use policies. Thus, we need a very simple commuting network algorithm able to generate the network of the European regions where we know little about the detailed commuting data.

Therefore, our algorithm considers each individual who wants to commute, from their living place to all the possible workplaces. It decides where they work following the classical rule inspired from the gravity law consisting in a compromise between the job offers and the distances to the jobs. The further away the job is, the more important the offer should be to be considered for the decision. Inversely, only the quantity of offers is important for the decision when these offers are close by. Each time a commuter has chosen its working place, the municipality counters for people living in and working outside on the one hand, and people working there and living outside on the other hand are decreased. As the initial values of these counters correspond to the value of the origin-destination table made available by the local statistical institute, the number of in-commuters and out-commuters of each municipality is respected without having to calibrate specific parameters.

The individual choice for a job location is probabilistic. Practically, the decision is stochastic. Thus, the model is stochastic as well. Each time we run it, we obtain a different network instead of obtaining an optimized network making deterministic the flow between the related municipalities. Especially for the latter, a deterministic approach does not appear relevant since the local choice is largely the fruit of a "by chance" process. The validation of the model shows that we obtain a good fit of the network given by the observed data. These results are very stable; the stochasticity of the model thus appears as a source of local diversity without perturbing the statistical properties of the network.

For the deterrence function, we decided, mainly inspired by the gravity law, to use a power law but another function could be tested.

This paper begins by a description of the characteristic of our four case study regions. Certainly, as part of a EU project, the study of different regions from a diversity of countries is done. They are defined as sets of NUTS3 regions for each country; these regions vary in size, population and other economic and social properties. We are interested in the inter-municipality commuting network. Very few papers deal with this on a small scale. Some analyse the inter-municipality commuting network (De Montis et al. 2007) (De Montis et al. 2010) showing that the Sardinian and the Sicilian inter-municipal commuting networks exhibit a traffic property based on a power law with exponent 2. Others, such as (Thorsen and Gitlesen 1998) compare different spatial interaction models, by an empirical evaluation of the municipalities of a Norwegian region. One analyses at the district level (which is higher than the municipality level) the German commuting network (Patuelli et al. 2007) using a comparison of two spatial interaction models. The set of publications shows the interest in such an approach to study the evolution of these type of networks over time.

After presenting the regions, we describe the proposed model and the very simple way we calibrated it. The third part is entirely dedicated to the validation of the model looking at different space-level properties.

## 2. Regional commuting network structures – Specific differences and global properties.

The first part of our study concerns the analysis of our study regions. The local statistical offices[3] provide all the information needed to characterize the structure of the commuting network. We consider: two separated NUTS2 regions in France (Auvergne and Bretagne), each composed of four NUTS3 regions; a group of two NUTS3

---

[2] PRototypical policy Impacts on Multifunctional Activities in rural municipalities – EU 7th Framework Research Programme; 2008-2011; https://prima.cemagref.fr/the-project

[3] In France: thanks to the Maurice Halbwach Center which made available the complete French origin-destination tables for commuters in 1999.

In Germany: Commuting data was purchased from the German Federal Employment Agency (Bundesagentur für Arbeit) for the year 2000.

In the United Kingdom: Origin-destination data was obtained via the Office for National Statisics NOMIS online database (https://www.nomisweb.co.uk/) for the year 2001.

regions in the UK (Nottinghamshire and Derbyshire); and a group of two NUTS3 regions in Germany (the Altmark region conformed by the districts of Stendal and Salzwedel).

The selected regions differ on many aspects: the number of municipalities, geographical structure, and socio-economic characteristics. They were chosen by the European project because they are all rural regions with different socio-economic situations. Table 1. presents some basic characteristics of each case study.

The objective of this first analysis is to determine the characteristics of the commuting networks composed by the regional commuting flows that are present in each region. For this analysis, we create a commuting matrix from a dataset containing the number of individuals that commute (i.e. reside in one settlement and work in another) within each of the selected regions. A representative section of the matrix used is shown in Table 2. Each row represents the place of residence and each column represents the working place; the cell at the intersection of each row and column contains the number of persons living and working in the corresponding row and column. For our analysis we ignore the cells in the matrix diagonal as they represent non-commuting individuals (persons living and working in the same place).

**Table 1. Characteristics of selected study regions**

| Region | Number of municipalities | Average size of a municipality (in number of inhabitants) | Average inter-municipality distance (in km) | Type of distance | Number of commuters living and working in the region | Part of the commuters living in the region and going to work outside of the region | Total surface (in km²) |
|---|---|---|---|---|---|---|---|
| Auvergne (France) | 1310 | 1024 | 88 | Euclidian | 261822 | 7.73 | 26.013 |
| Bretagne (France) | 1269 | 2447 | 99 | Euclidian | 608587 | 7.32 | 27.208 |
| Altmark (Germany) – subregions | 91 | 2527 | 50 | Shorter road | 16770 | **66.82** | 4.715 |
| Nottinghamshire /Derbyshire (UK) | 372 | 5300 | 44 | Shorter road | 573022 | 12.4 | 4.839 |

**Table 2. Example of commuting data from the Altmark Region**

|  |  | Municipality of the workplace | | | | | | |
|---|---|---|---|---|---|---|---|---|
|  |  | 81026 | 81030 | 81035 | 81045 | 81080 | 81095 | … |
| Municipality of residence | 81026 | 0 | 0 | 0 | 0 | 0 | 0 | … |
|  | 81030 | 0 | 0 | 0 | 3 | 0 | 0 | … |
|  | 81035 | 0 | 0 | 0 | 2 | 0 | 0 | … |
|  | 81045 | 0 | 2 | 2 | 0 | 2 | 2 | … |
|  | 81080 | 0 | 0 | 0 | 0 | 0 | 0 | … |
|  | 81095 | 0 | 2 | 0 | 8 | 0 | 0 | … |
|  | … | … | … | … | … | … | … | 0 |

After analyzing some global properties of the network structure we can observe that the presented regions have quite dissimilar behaviours.

The first global property of the network that we analyse concerns the distributions of the degrees. The degree is a property of the associated un-weighted network. For the construction of the un-weighted network we consider all the municipalities and we add a directed link between the municipality $i$ and the municipality $j$ if at least one individuals commutes from $i$ to $j$. The in-degree of a municipality $i$ ($k_{in}(i)$) is the number of links starting from $i$, while the out-degree ($k_{out}(i)$) is the number of links entering $i$.

The probability distributions of the in and out degrees are represented in figure 1.

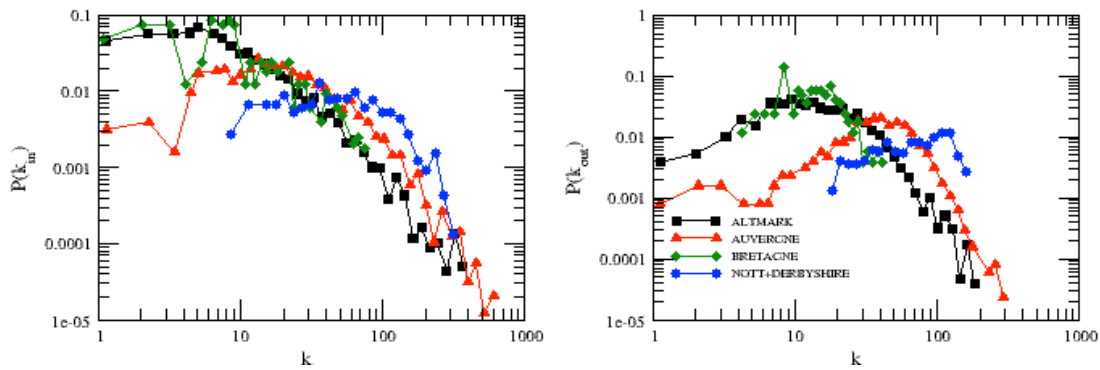

**Figure 1. In and Out degree distributions of each case study region**

As we can observe, the different case studies are characterized by very different behaviours according to the degree distribution. If we focus, for example on the out-degree distribution we can observe that the municipalities in the UK region always have quite a large and uniform degree distribution. It can be explained by the fact that in this case the number of commuters is extremely large, and the network is very dense in terms of links. This kind of uniform structure can be connected to the lack of working "hubs" able to attract workers more strongly than the other municipalities. This coincides with what we observe in the in-degree distribution where we see that few municipalities have a small in-degree while a considerable part has a high in-degree.

The situation in Auvergne and Bretagne, where the in-degree distributions suggest the presence of real "hubs" of the commuting network (a small but not unimportant part of municipalities reached much more than the others) is totally different.

We can finally observe that for Altmark, the total number of connections is generally lower, probably due to the fact that this region represents only a part of a larger commuting network. Indeed in the case of the Altmark region (former Eastern Germany territory), which is located near the state of Niedersachsen (former West Germany territory), the proportion of commuters is more than double that of the state of Sachsen-Anhalt.[4]

Another important consideration concerns the distribution of the distances covered by the commuters. This measure is presented in Figure 2.

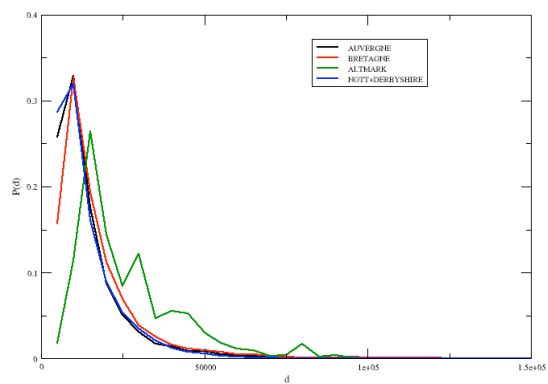

**Figure 2. Distribution of the commuting distances (in meters) for the selected case studies**

The distribution of the distances shows that in the UK regions smaller distances are favoured. This confirms the intuition that in the region almost all the municipalities are at the same level in terms of job offers (no hubs), so it is not needed to travel long distances to find a job. We notice that it is exactly the opposite in the Altmark case, where a significant share of the commuters can travel up to 80 km.

---

[4] Source: Statistisches Landesamt, calculations made by Konjunkturteam Altmark (http://www.stendal.hs-magdeburg.de/project/konjunktur/risa/altmark/frame_altmark_sv_beschaeftigtencharts.htm)

In this section we focused on the fact that the selected case studies present strongly dissimilar behaviours regarding the global properties of the commuting networks.

In the following section we focus on the choices of the single commuters and we try to see if a common base, centred on the individuals' behaviours, can be retrieved under all these dissimilarities.

## 3. A Monte-Carlo simulation approach to generate realistic commuting networks

The usual methods for reconstructing the structure of commuting networks are based on the gravity law function explained in the previous section. The main issue with this approach is that it is not always easy to calibrate the three parameters of the gravity law model (Williams 1976). We propose a simple model that presents a higher level of universality and which can be applied with a good degree of confidence to all the case study regions.

### 31. The individual-level generation model

The model is based on the individual choices of the commuters, namely the surplus of people, in the active class, that are not able to find an occupation in the place where they live.

When looking for an occupation outside of the living place, two factors can influence the choice of the destination: the distance of the possible workplace and its "attractiveness" (which is defined by the number of jobs it offers). The further away the possible destination is, the more its attractiveness will matter in the decision. If the possible destination is near, the settlement attractiveness becomes less significant for the individual's decision for a workplace.

We start from a typology of data that is usually available, for each municipality, in each case study:
- the total number of out-commuters ($R_i$), also called the job demand of the municipality $j$
- the total number of in-commuters ($Q_i$), also called the job offer of the municipality $i$
- the distances among each couple of municipalities ($d_{ij}$)

The algorithm associates to each commuter, in each municipality $i$, a working destination $j$ according to the job offers of the other municipalities in the region and the distance among the municipality $i$ and all the possible destinations.

The algorithm for the generation of the network evolves according to the following steps:

For each remaining commuter who has not already found a place of work, we:
1. Select a living municipality $i$ at random among the municipalities where it remains at least one out-commuter ($R_i>0$)
2. Select the working destination $j$ randomly following the probability distribution given by:

$$p_{i \to j} = \frac{Q_j d_{ij}^{-\beta}}{\sum_{j \neq i} Q_j d_{ij}^{-\beta}}$$

3. Update the number of out-commuters of $i$ and the number of in-commuters of $j$: $R_i=R_i-1$ . $Q_j=Q_j-1$
4. Recalculate the $p_{i \to j}$ distribution

The interaction between the two factors is characterized in the model with the parameter $\beta$ which captures the relative impact of the distance.

Using this algorithm we ensure that the generated network respects exactly the incoming and outgoing traffic from each node.

Different values of the parameter $\beta$ produce different distance and degree distributions for the generated networks. We calibrate the parameter, for the case studies where the complete information in the network is known, in order to have the same distance distribution as the one observed for the real network.

Analysing the calibration on the regions where the data is available, we can observe that with an appropriate choice of the parameter *β*, we are able to reconstruct every commuting network structure accurately.

The calibration procedure and the analysis of the accuracy of the generation algorithm are presented in the next paragraph.

## 32. Model calibration

The proposed model depends on a spatial parameter $\beta$ which represents the relative importance of the distance to the destination in the choice of the working place.

A typical fingerprint that distinguishes the commuting networks is the distribution of the travelled distance for each worker. We use this information to calibrate the parameter *β*. In fact, each value of $\beta$ produces a network with a typical distance distribution, as it is displayed in Figure 3 for the Auvergne case study.

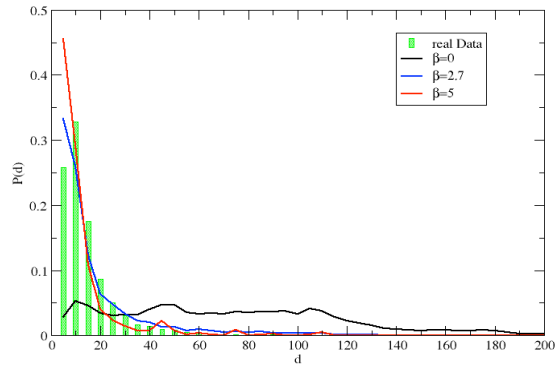

**Figure 3. Distance d in km distribution for the real network and three different *β* values for the Auvergne case study**

We observe that, for excessively low values of $\beta$, the preference toward distant working places is overestimated, while for excessively high values the choice for close places is overestimated. We calibrate $\beta$ in order to minimise a distance between the generated travelled distance distribution and the one building from the Census data (called observed data). The minimised distance is the Kolmogorov-Smirnov one:

$$D_{KS} = \sup_{d} \left| P^c{}_o(d) - P^c{}_g(d) \right|$$

where $P^c{}_{o/g}(d)$ are the cumulative distance distributions for the observed (*o*) and generated (*g*) networks.

For each case study we calculate this distance for different values of $\beta$ and we choose the minimum of the function $D_{KS}(\beta)$ as the calibration value. This process is shown by the figure 4.

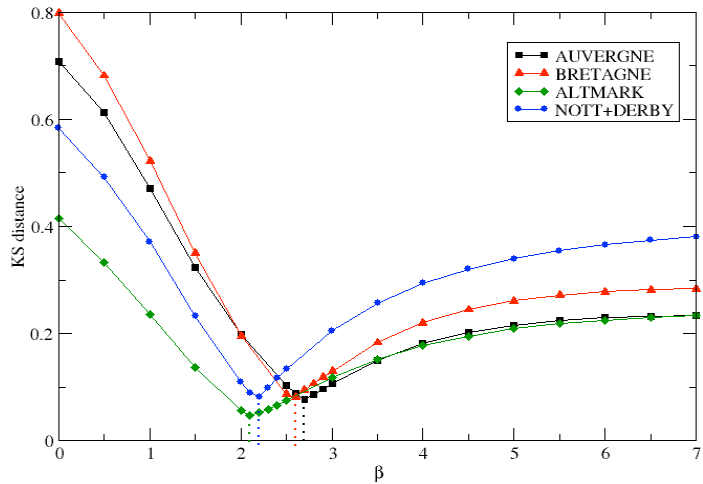

**Figure 4. Calibration process for the four case study regions based of the minimisation of the average Kolmogorov-Smirnov distance**

Since the model is stochastic, the Kolmogorov-Smirnov distance (KS) is averaging on the KS measured on the 100 replicates we simulate for each $\beta$ value. The variation of the measured KS is very low: it is atmost 1.13% of the average KS.

## 33. The validation

To assess the quality of the generated network, we compare its properties to the properties of the observed network. Two different kinds of properties are investigated: a first group measured on the municipality network where we consider that two municipalities are linked when at least one worker commutes between them, whatever the origin-destination is; a second one measured on the weighted network which has direct links weighted by the number of individual commuting from a given municipality to another given one.

The first set considers two different indicators:
- the ability of the generated data to fit the observed in and out degree distributions of the "municipality" network;
- the traffic density distribution describing the density of each weight that can be associated to a undirected link. For an arc between two municipalities, this weight is the sum of the individuals going from one municipality to the other in both directions through the arc.

The second set measures the common part of commuters to both the generated and the observed network in the weighted directed network.

These two statistics have not been used to generate simulated networks. Moreover, we must remember that the number of people looking for a job in a municipality $i$ ($R_i$) and the job offers in a municipality $j$ ($Q_j$) are exactly reproduced by the generation algorithm whatever the municipalities.

## 33.2 The properties of the municipality network

In this study we focus on two quantities that describe respectively the topological properties of the network and the flows' characteristics: the in and out degree distribution (p($k_{in}$) and p($k_{out}$)) and the traffic distribution (p(T)).

First of all, it is worth noticing that these indicators are influenced by the choice of the parameter $\beta$. As we can observe in Figure 5 for the Auvergne case study, for $\beta$=0, i.e. when the geography is not important at all, higher degrees and lower traffics are observed. As the geography becomes more important the maximum degree decreases and the maximum traffic increases. It is due to the fact that when geography is not important people choose their working destination in a wider range of available municipalities, while a strong geographical constraint forces to choose only between the nearby municipalities. As a consequence of this, the traffic on this smaller number of connections will also be globally higher.

A crucial point can be observed in Figure 6 concerning the comparison with the observed data. As we can notice, even if we did not use these explicit measures for the calibration, the distributions at the calibration point ($\beta$=2.7) fit the distributions of the observation network perfectly.

Figure 6 shows the comparison between these measures for the observed network and the generated ones for the other case studies.

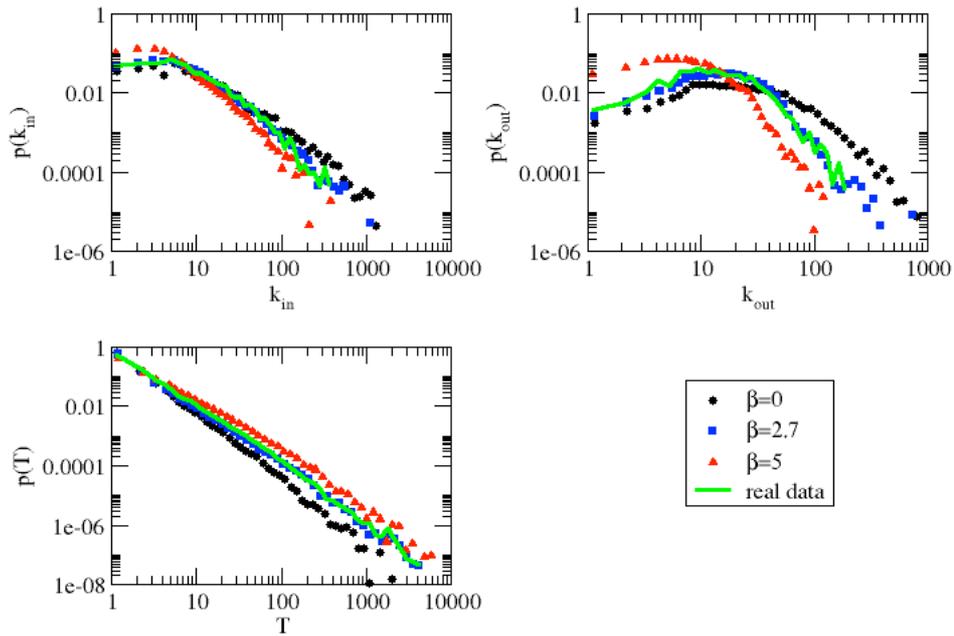

**Figure 5. In and out degree distribution and traffic distribution for different values of $\beta$ and for the real network for the Auvergne case study. The results for the generated networks are averaged on 100 realizations of the model.**

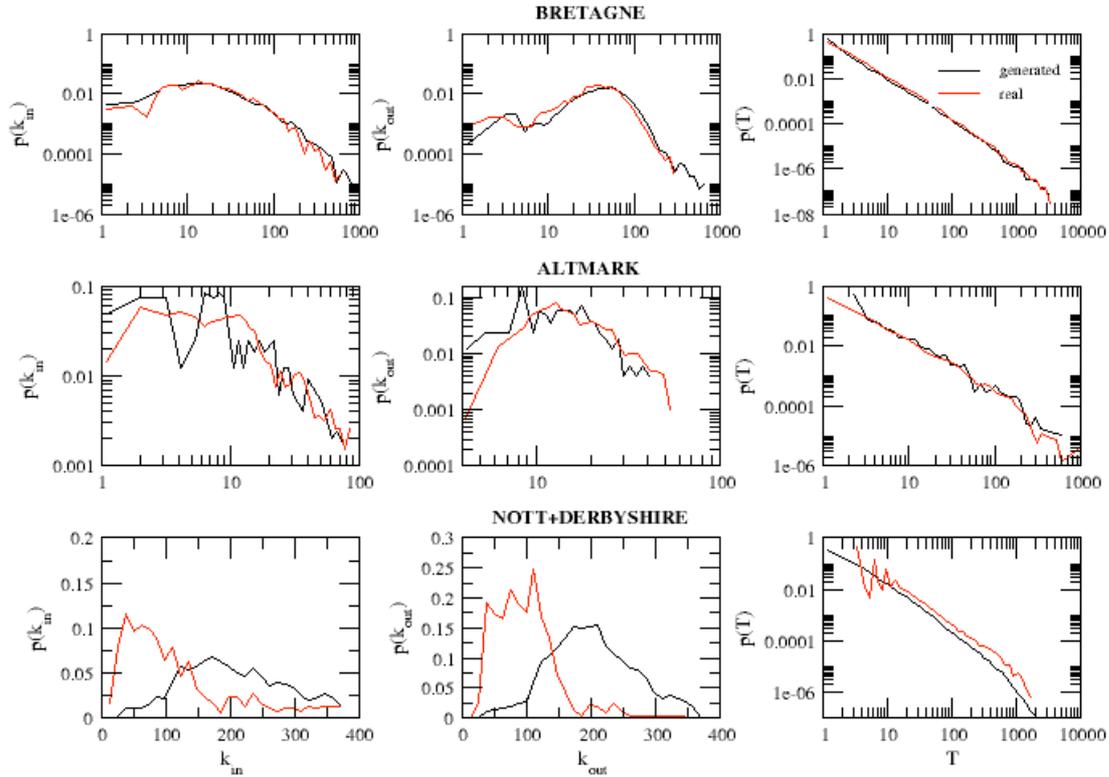

**Figure 6. In and out degree distribution and traffic distribution for the generated networks at the calibration point and for the real network for the Bretagne, Altmark and UK case studies. The results for the generated networks are averaged on 100 realizations of the model.**

As we can notice in Figure 6 the traffic distribution is well reproduced in all the case studies. It is not the case for the degree distributions where, in the UK case study, the generation process completely fails in the estimation. This discrepancy is not due to the algorithm but mainly due to the Census data. Indeed, in UK, a small-cell adjustment method (Stillwell and Duke-Williams 2007) is applied to prevent disclosure of data. In particular, this method suppresses the flow value 1 and 2 replacing them by 0 or 3. This adjustment makes the definition of a link between two municipalities different in the model beyond the data and in the generated network through our algorithm. In the data, two municipalities are linked only if at least three individuals commute among them. In our generated network, they are linked if at least one individual commute among them. A large number of municipality couples are in the reality linked by only one or two individuals and these couples are underestimated in the corrected UK data. That is the reason why our generated network seems to overestimate the connectivity between municipalities.

## 33.1 The common part of commuters

We need an indicator to compare the simulated commuting network and the observed commuting network. Let a $R \in M_n(N)$ be commuting network when $R_{ij}$ is the number of commuters from the municipality $i$ to the municipality $j$. Let $S \in M_n(N)$ be another commuting network between same municipalities. We can compute the number of common commuters (NCC) between $R$ and $S$ (1) and the number of commuters (NC) in $R$ (2). We finally compute the common part of commuters (CPC) (3) which seems to us a good indicator of the prediction quality. This indicator can be seen as a variant of the Sørensen index which is simplified due to the fact that the two compared matrices have the same size. This error has a more obvious interpretation than a more classical error, as the absolute and quadratic distances.

$$NCC_n(S,R) = \sum_{i=1}^{n}\sum_{j=1}^{n} \left( S_{ij} 1_{(R_{ij}-S_{ij}) \geq 0} + R_{ij} 1_{(R_{ij}-S_{ij}) < 0} \right) \qquad 1$$

$$NC_n(R) = \sum_{i=1}^{n}\sum_{j=1}^{n} R_{ij} \qquad 2$$

$$CPC = \frac{NCC_n(S,R)}{NC_n(R)} \qquad 3$$

This gives us an indicator for one replicate if we compare the generated distribution for one replicate to the observed one. We do the same with all the 100 replicates for a given $\beta$ value and compute the average of the 100 CPC obtained to evaluate the quality of the model. The CPC varies at most by 1.76% of the average. Then the stochastic model is very stable. Table 3 shows the average CPC for each case study regions.

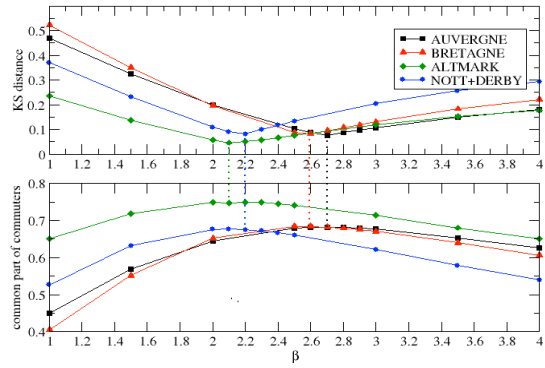

**Figure 7. Common part of commuters for different $\beta$ values for each case study region (compared to the calibration graph in Figure 4)**

Figure 7 presents the average CPC for each region and for different $\beta$ values. It is noticeable that the best value of the average CPC function is very close to the one given by the calibration value of $\beta$ for all the studied regions. This point is stressed by the dotted line showing where the minimum of the KS distance is situated on the common part of commuters function. The proximity in terms of $\beta$ of the minimum of the KS distance function and the maximum of the common part of commuters function is quite surprising and looks like a good quality indicator. We can also notice that, both the KS distance and the common part of commuters varies weakly between $\beta = 2$ and $\beta = 3$. Moreover the $\beta$ value seems to vary in the same way that the acreage or the average distance between the municipality of the region for which the Altmark and the Nottinghamshire and the Derbyshire are close with a small value and the Auvergne and the Bretagne regions are close with a large value (see table 1).

**Table 3. Average Common Part of Commuters for the four case study regions**

| Region | $\beta$ | Average Common Part of Commuters |
|---|---|---|
| Auvergne | 2.71 | 0.683 |
| Bretagne | 2.59 | 0.684 |
| Altmark | 2.1 | 0.751 |
| Nottinghamshire and Derbyshire | 2.2 | 0.676 |

## 4. Discussion and conclusions

We propose a very simple stochastic individual-based model able to generate a commuting network with good accuracy. This model is based on the doubly-constrained model proposed by Wilson in 1970. It is built on the same principles: an individual tends to choose a job location depending on the job offers and the distance to the offer. The distance's effect decreases when the distance increases following a function that we have chosen as a power law. Our model has only one parameter, in contrast with the doubly-constrained model that has three parameters. Our algorithm ensures that the number of out-commuters and in-commuters for each municipality is respected. Thus the two balancing factors of the doubly-constrained model dedicated to ensure the same equality can be suppressed.

The algorithm is validated on four case-study regions situated in France, Germany and the United-Kingdom. We compare the properties of the observed network given by the complete origin-destination table to those of the generated networks. We conclude that the in and out degree distributions of the municipality network, the traffic distribution of the same network are well fitted by the generated networks' distributions. Moreover, the common part of commuters of the weighted directed network to the observed and the generated networks appears quite high for all the case study regions. Incidentally, we have noticed that the parameter, calibrated on the commuting distribution, has an optimum value very close to the parameter value giving the larger part of commuters.

In this discussion, we come back to our first objective. We wanted to find a very simple commuting generation model allowing us to generate a statistically representative commuting network of a region for which only the quantity of in-commuters and the quantity of out-commuters is available at the municipality level.

The proposed model appears quite relevant. However, the aggregated statistics generally available at the municipality level correspond to all the commuters-in and all the commuters-out of the municipality. A part of these commuters can respectively live or work outside the region. To be sure that our model gives some significant network, it has to be applied on a region where these commuters linked to the outside represent a non significant part of the total number of commuters. In other words, the region should be what Paelink and Nijkamp called a "polarized region" in 1975: "*a connex area in which the internal economic relationships are more intensive than the relationships with respect to regions outside the area*" (Cörvers et al. 2009; Konjar et al. 2010) . It is possible to conclude, through aggregated data at the regional level, or expertise, if a region is or isn't sufficiently independent from another regarding the labour market.

The second point to discuss regarding the relevance of our model for our objective concerns its calibration. Most of the known power-law network has an exponent value situated between 2 and 3. Our first case studies seem to show that the exponent of our power-law deterrence function varies in the same range. We notice that the error remains quite low between these two boundaries for $\beta$. It also seems that the better $\beta$ optimal value can be chosen looking at the acreage or the average inter-municipality distance of the region.

Finally, concluding on the way to calibrate our model or on its quality requires a larger number of case study regions having a lot of geographical and socio-economic differences.